\documentclass[prl,aps,twocolumn,groupedaddress,showpacs,floatfix,preprintnumbers]{revtex4}
\usepackage{amsmath,amssymb,multirow,epsfig,bm,dcolumn}

\newcommand{\beq}{\begin{equation}}
\newcommand{\eeq}{\end{equation}}
\newcommand{\bea}{\begin{eqnarray}}
\newcommand{\eea}{\end{eqnarray}}

\begin{document}
\preprint{NT@UW-15-14,  PNNL-SA-114173, LA-UR-15-28092}

\title{Induced Fission of $^{240}$Pu within a Real-Time Microscopic Framework}
\author{Aurel Bulgac$^1$}
\author{Piotr Magierski$^{2,1}$}
\author{Kenneth J. Roche$^{3,1}$}
\author{Ionel Stetcu$^4$}

\affiliation{$^1$Department of Physics, University of Washington, Seattle, WA 98195--1560, USA}
\affiliation{$^2$Faculty of Physics, Warsaw University of Technology,
ulica Koszykowa 75, 00-662 Warsaw, POLAND}
\affiliation{$^3$Pacific Northwest National Laboratory, Richland, WA 99352, USA}
\affiliation{$^4$Theoretical Division, Los Alamos National Laboratory, Los Alamos, NM 87545, USA}

\begin{abstract}

  We describe the fissioning dynamics of $^{240}$Pu from a
  configuration in the proximity of the outer fission barrier to full
  scission and the formation of the fragments within an implementation
  of the Density Functional Theory (DFT) extended to superfluid
  systems and real-time dynamics. The fission fragments emerge
  with properties similar to those determined experimentally, while
  the fission dynamics appears to be quite complex, with many excited shape
  and pairing modes. The evolution is found to be much slower than
  previously expected and the ultimate role of the collective inertia is 
  found to be negligible in this fully non-adiabatic treatment of nuclear 
  dynamics, where all collective degrees of freedom (CDOF) are included (unlike 
  adiabatic treatments with small number of CDOF).
  \end{abstract}

\date{\today}



\maketitle

Nuclear fission has almost reached the venerable age of 80
years~\cite{Natur_1939r,Nature_1939r} and it still lacks an
understanding in terms of a fully quantum microscopic approach. This
is in sharp contrast with the theory of superconductivity, another
remarkable quantum many-body phenomenon, which required less than half
a century since its discovery in 1911~\cite{KNAV_1911r} until the
unraveling of its microscopic mechanism in 1957~\cite{PhysRev_1957r}.
N. Bohr~\cite{Nature_1939r1,PhysRev_1939r, PhysRev_1953r,
  PhysRev_1952r} realized that the impinging low energy neutrons on
uranium targets leading to the nuclear fission proceed through the
formation of a very complex quantum state, the compound nucleus, which
has a very long life-time. In a compound state the initial simple wave
function of the impinging neutron is fragmented into a wave function
of the nucleon+nucleus system with approximately one million
components, as level density suggests~\cite{PhysRevC_2005r}. In this
respect this is similar to a particle in a box with a very small
opening, consistent with the long lifetime of a compound nucleus
state.  Eventually, due to the interplay of the Coulomb repulsion
between the protons and the nuclear surface tension, the nuclear shape
evolves like a liquid charged drop and the compound nucleus reaches
the scission configuration, leading predominantly to two emerging
daughter nuclei. It was a great surprise when in the 1960s it was
realized that the independent particle model proved to play a major
role in the fission dynamics.  At that time it became clear that
independent particle motion of nucleons and shell effects play a
remarkable role and lead to a very complex structure of the fission
barrier~\cite{RMP_1972r,RMP_1980r} and to a potential energy surface
much more complicated than that suggested by a liquid drop model
considered until then.  On its way to the scission configuration a
nucleus has to overcome not one, but two - the double-humped fission
barrier - and sometimes even three potential
barriers~\cite{RMP_1972r,RMP_1980r}.  As in low energy neutron induced
fission the excitation energy of the mother nucleus is relatively
small, the compound nucleus has a very slow shape evolution and it was
reasonable to assume that the shape evolution is either damped or
over-damped. And since the presence of shape isomers has been
unequivocally demonstrated, experimentally and theoretically, the
dominant phenomenological approach to fission dynamics based on
compound nucleus ideas, liquid drop, shell-corrections, and the role
of fluctuations described within a Langevin and statistical
approaches~\cite{PhysRevC_1976r, NuclPhys_1992r, PhysRevLett_2011r,
  PhysRevC_2011r, PhysRevC_2013r5, PhysRevC_2014r, PhysRevC_2015r, 
  arxiv_2015r3, PhysRevC_1976r5, PhysRevC_2015r5} has been born.
  
It became clear over the years that the fermion pairing and
superfluidity play a critical role in nuclear fission, though in a
vastly different manner than in the case of
superconductivity~\cite{NuclPhys_1990r, NuclPhys_1994r}.  Pairing
correlations (either vibrations or rotations) are ubiquitous in
nuclei~\cite{PhysRev_1958r} and they are expected to play a leading
role in the nuclear shape dynamics~\cite{NuclPhys_1990r,
  NuclPhys_1994r, PhysRevLett_1997r, PhysRevC_1978r}.  The shape
evolution of nuclei appears somewhat surprising at first sight, since
typically a nucleus is stiffer for small deformations and rather soft
for large deformations. Hill and Wheeler~\cite{PhysRev_1953r} had the
first insight into the origin of this aspect of nuclear large
amplitude collective motion: the jumping from one to another diabatic
potential energy surface and the role of Landau-Zener transitions.
The most efficient microscopic mechanism for shape changes is related
to the pairing interaction.  The difficulty of making a nucleus
fission in absence of superfluidity was illustrated within an
imaginary time-dependent Hartree-Fock approach treatment (instanton in
quantum field theory parlance) of the fission of $^{32}$S into two
$^{16}$O nuclei~\cite{NuclPhys_1989r}. The initial and final states
have an obvious axial symmetry, with occupied single-particle
$m$-quantum states $\pm1/2^5,\pm3/2^2,\pm5/2^1$ and
$\pm1/2^6,\pm3/2^2$ for protons and neutrons, respectively in the
mother and daughter nuclei, where the superscript indicates the
number of particles with the corresponding $m$-quantum number.  In the
absence of short-range pairing interactions, particularly effective at
connecting time-reversed nucleon pair states $(m,-m)$ with $(m',-m')$, and in
particular the transition $(5/2,-5/2) \rightarrow (1/2,-1/2)$ in
$^{32}$S, fission is possible only if an axially broken symmetry
intermediate state is allowed.

Since the late 1970s~\cite{PhysRevC_1978r} and in particular during
the last decade an alternative approach in the theoretical treatment
of fission dynamics started gaining ground with the implementation of
the philosophy of the DFT~\cite{hk, ks, wk, monograph1, monograph2,
  monograph3, tddft, monograph4, oliveira} and its various
modifications~\cite{ arxiv_1511r, PhysRevC_2014r11, 
  PhysRevC_2015r1, PhysRevC_2015r2, arxiv_2015r,
  arxiv_2015r1, PhysRevC_2009r, PhysRevC_2011r1, PhysRevC_2014r1,
  PhysRevC_2014r2, PhysRevC_2014r3, PhysRevC_2008r, Maruhn}.  DFT is viewed as
an alternative to solving the Schr\"odinger equation, in which the
role of the many-body wave function is replaced by the one-body
density matrix. DFT, however, does not provide a constructive recipe
to determining the underlying functional. Application to nuclear
physics requires a generalization of the most successful DFT
implementation: the Kohn-Sham Local Density Approximation
(LDA)~\cite{ks} to fermionic superfluid and time-dependent phenomena
-- the Superfluid LDA (SLDA) and its time-dependent (TD) extension, a
formalism based on local meanfield and pairing potentials.  DFT is
formulated by construction to appear as the
Hartree-Fock or Hartree-Fock-Bogoliubov approximation (sometimes referred 
improperly in our opinion as HF or HFB), but it is in
principle, though not in practice, exact.  With the use of Quantum
Monte Carlo results for cold atoms and 
phenomenological input for nuclear systems, (TD)SLDA has
been validated against a wide range of experimental results ~\cite{PRL__2002, PRC__2002, PRL__2003,
  PRL__2003a, PRL__2003b, PRA__2007, PRL__2008, PRL__2009,
  Science__2011, LNP__2012, PRL__2012, ARNPS__2013, PRL__2014,
  PRA__2015, PRC__2011, PRL__2015}.
  
The structure of the nuclear energy density functional (NEDF) is still
largely based on phenomenology~\cite{arxiv_2015nedf} and our approach
here is based on the popular Skyrme parametrization
SLy4~\cite{NuclPhys_1998} and the SLDA treatment of the pairing
correlations~\cite{PRL__2003a}. The numerical aspects of our approach
have been described in great detail in Refs.~\cite{Science__2011,
  PRL__2015, JPhys_2008, PRC_2013a} and the results presented below
have been obtained by solving the TDSLDA equations for a $^{240}$Pu
nucleus in a simulation box $22.5^2\times40$ fm$^3$, with a lattice
constant corresponding to a relatively high momentum cutoff $p_c \approx 500$ MeV/c,
and with no spatial restrictions.  The time step used was 0.119 fm/c for
up to 120,000 time steps, using  $\approx 1,760$ graphic processing units (GPUs) for
a total wall time of 550 minutes.  The TDSLDA equations, which
amounted to $\approx 56,000$ complex coupled nonlinear TD
3D partial differential equations, were solved using a
highly efficient parallelized GPU code~\cite{PRL__2014, PRA__2015,
  PRL__2015} on Titan~\cite{cray}.

\begin{figure}
\includegraphics[height=9cm]{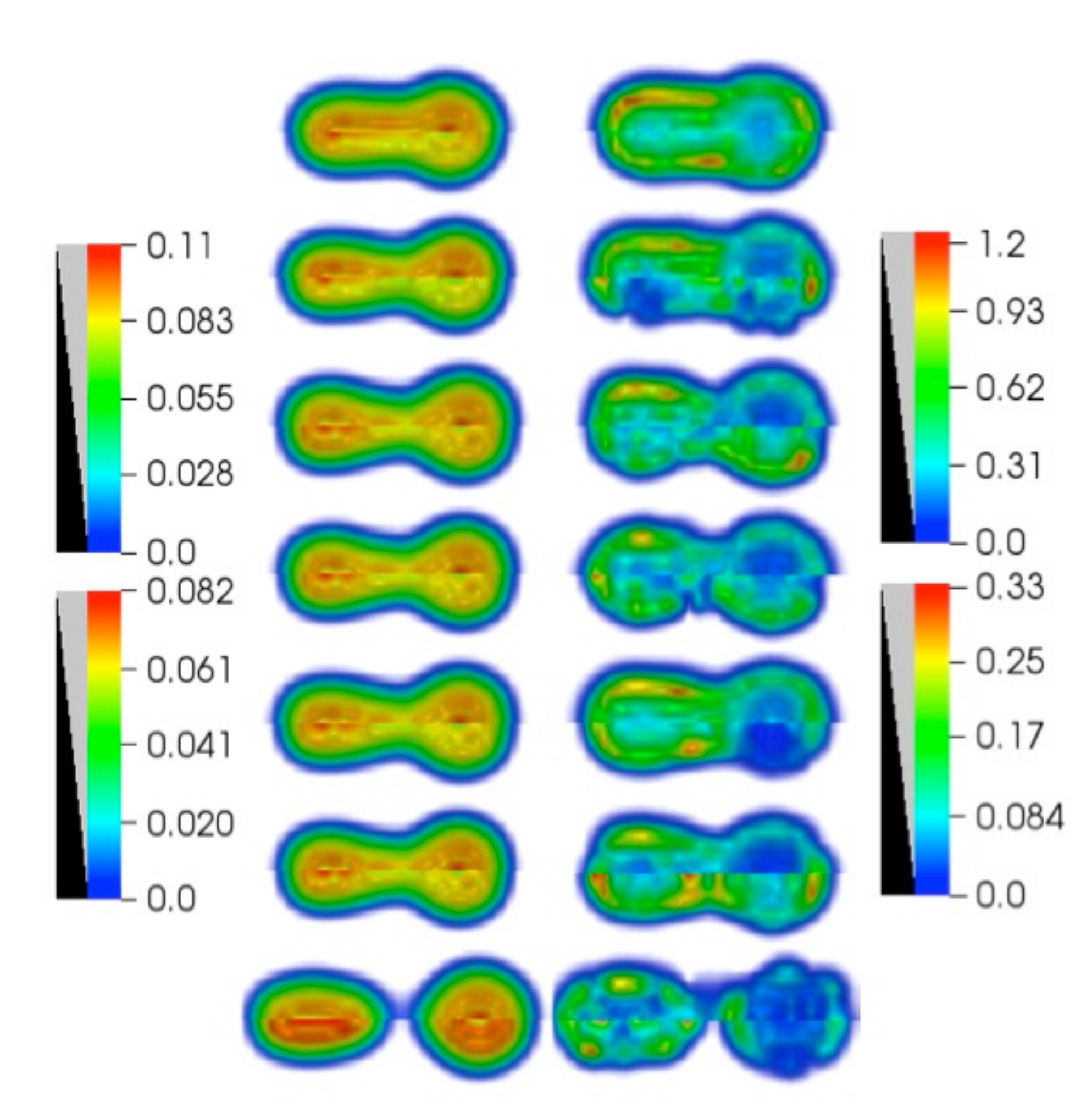}
\caption{ (Color online) \label{fig:Pu240} The left column shows 
the neutron/proton densities in the top/bottom half of each frame.
In the right column the pairing field for the neutron/proton systems
are displayed in the top/bottom of each frame respectively. The
time difference between frames is $\Delta t = 1600$ fm/c. The range of values 
are (0, 0.1) and (0,0.07) fm$^{-3}$ for $\rho_{n,p}({\bm r})$ and  
(0, 0.9) and (0, 0.7) MeV for $\Delta_{n,p}({\bm r})$  
respectively, with colorbars on the left/right for densities/pairing 
gaps, with upper/lower ones for neutrons/protons respectively. 
These frames are equally spaced in time for the case of the simulation S1, 
see Table~\ref{table:A}. }
\end{figure}

\begin{figure}
\includegraphics[width=8.1 cm]{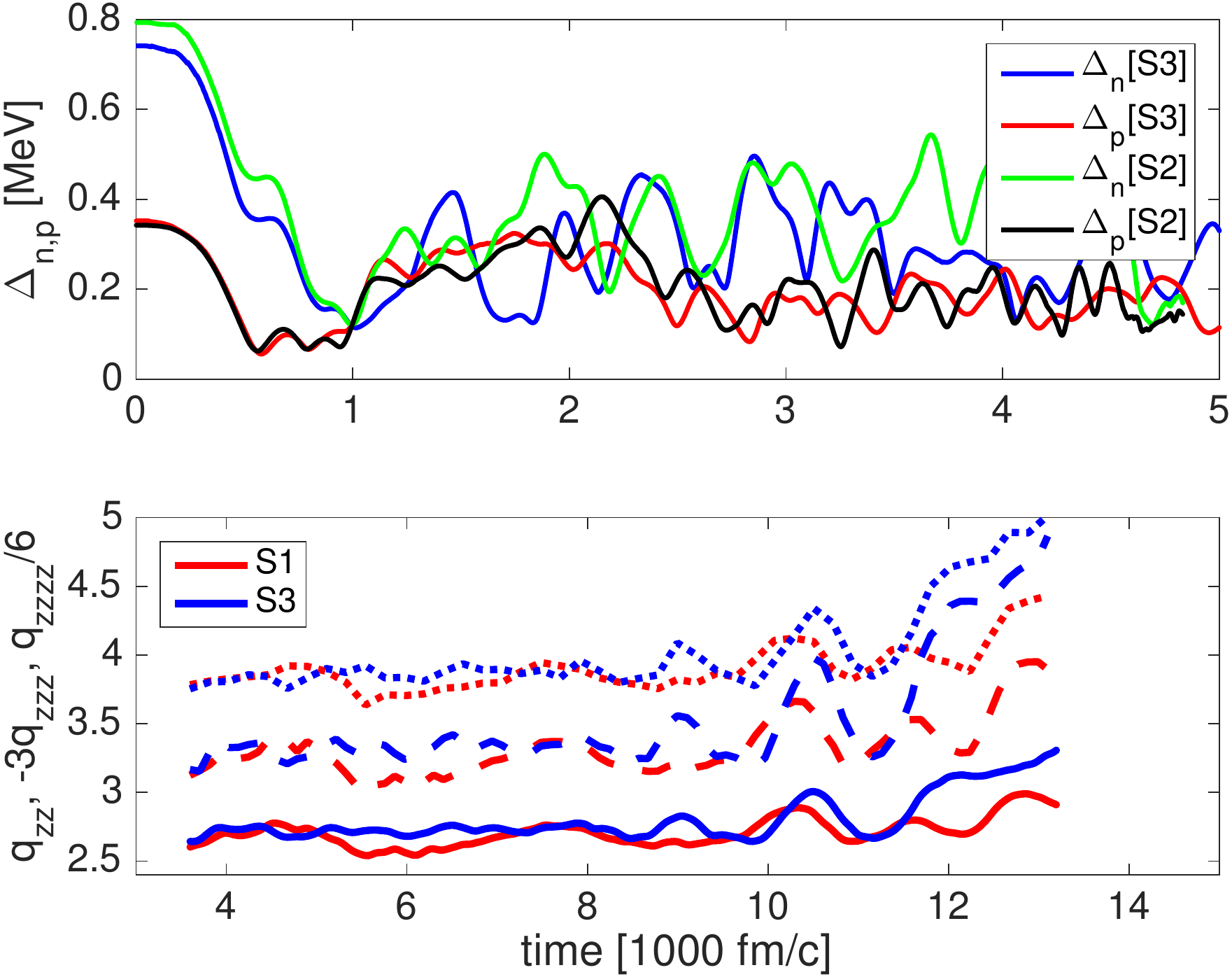}
\caption{ (Color online) \label{fig:Delta}  The time dependence 
of spatially averaged $|\Delta_{n,p}({\bf r,t})|$  
for S2 (mixed pairing) and S3 (volume pairing)  in the 
upper panel and in the lower panel the scaled mass moments 
$q_{20}(t) = \int d^3(3z^2-r^2)/A^{5/3}\rho(\bm{r},t),\quad 
  q_{30}(t) = \int d^3 z(5z^2-3r^2)\rho(\bm{r},t)/A^2, \quad
  q_{40}(t) = \int d^3 (35z^4-30z^2r^2+3r^4)\rho(\bm{r},t)/A^{7/3}$
with solid, dotted, and 
dashed lines respectively, for S1 (red) and S3 (blue) [fm$^L]$, see Table~\ref{table:A}. }
\end{figure}

A $^{239}$Pu nucleus bombarded with low-energy neutrons needs a very
long time to evolve from its initial ground state shape until it
reaches the outer fission barrier.  In a constrained self-consistent
calculation we bring the nucleus to a shape and an energy in the
immediate proximity of the outer fission barrier (at zero temperature).  Starting from this
configuration we follow the nuclear dynamics within the TDSLDA
approach until the two fragments are clearly separated, see
Fig.~\ref{fig:Pu240}.  A summary of our results is presented in
Table~\ref{table:A} and complemented with movies of the real-time
simulations~\cite{movies}.  The main difference between various
simulations is in the character of the pairing correlations. Over the
years several distinct parameterizations of the pairing coupling
constant(s) have been suggested~\cite{PRC_2009a}, basically various
mixtures of the so-called volume and surface pairing, as compelling
\textit{ab initio} information is still lacking. The isospin symmetric
density-dependent pairing coupling constant is
$
g_{\tt eff} ({\bf r}) =g\left (1
-\eta\frac{\rho(\bf{r})}{\rho_0}\right ), \label{eq:eta} 
$
where $\rho(\bf{r})$ and $\rho_0$ are the total and the saturation
nuclear densities.  The extensive phenomenological information
gathered so far for ground states of nuclei fails to point to a well
defined value of the parameter
$\eta$~\cite{PRL__2003a,PRC_2009a}. The dynamics, as we 
demonstrate, depends strongly and non-monotonically on the parameter 
$\eta$. Fission dynamics requires a very
efficient mechanism for the shape evolution, which is directly linked
to transitions of the type $(m,-m)\rightarrow (m',-m')$, for which the
pairing interaction is particularly effective~\cite{NuclPhys_1990r,
  NuclPhys_1994r, PhysRevLett_1997r}.  The frozen occupation
probabilities approximation~\cite{arxiv_2015r, arxiv_2015r1} used in
the past, as well as a naive TDHF treatment \cite{Maruhn}, fail in this respect, as
they do not allow the needed transitions, from levels with high
$m$-values in the mother nucleus to levels with low $m$-values in the
daughter nuclei, to take place~\cite{PhysRevLett_1997r} and a nucleus
very often fails to fission or requires an inordinate amount of
push~\cite{PhysRevC_2014r11, PhysRevC_2015r2,arxiv_2015r, arxiv_2015r1}.  In some cases
the axial symmetry beyond the outer barrier could be broken, see
e.g. Ref,~\cite{PhysRevC_2015r1}, and a suitable valley exists in the
potential energy surface and fission can proceed.  The approximate
treatment of the pairing correlations within the TDBCS
approximation~\cite{PhysRevC_2015r, PhysRevC_2015r1} violates the
continuity equation~\cite{PhysRevC_2012r}. There is no question that a
smooth transfer of the nuclear matter from the waist of the mother
nucleus, which allows the nucleus to elongate and eventually to lead
to the neck formation, is expected for any approach to fission
dynamics.  The TDSLDA is so far the only theoretical framework with an
NEDF that satisfies all expected symmetries and theoretical
constraints.  At the same time in SLDA solutions all symmetries can
be broken, a situation similar to ferromagnets described by the
Heisenberg Hamiltonian.  In TDSLDA the evolution is by default smooth
and various contributions to the energy (often referred in literature
as collective potential and kinetic energies) are always continuous as
a function of the nuclear shape, unlike traditional approaches,  
Refs.~\cite{PhysRevC_2008r, PhysRevC_2014r1}. TDSLDA eschews
the need to evaluate the inertia tensor, to introduce or guess the
collective coordinates, or invoke the adiabaticity of the evolution.
The pairing field often reaches very small values, a situation 
also encountered in the study of the Higgs pairing mode 
\cite{ PhysRev_1963r4, NatPhys_2015} in other systems 
\cite{PRL__2009, PhysRevLett_2004r,  arxiv_2007r4, 
PhysRevLett_2006r4, PhysRevA_2015r},  when 
the pairing field can attain even exponentially small values  
for long periods of time, only to revive again. As 
in case of density oscillations~\cite{PRC__2011, PRL__2015}
(studied within the Random Phase Amplitude limit), TDSLDA 
describes correctly the pairing vibrations, in case 
vanishing static pairing.

\begin{table*}[th]

  \caption{ \label{table:A} 
    The simulation number, the pairing parameter $\eta$, the excitation energy
    ($E^*$) of $^{240}_{94}$Pu$_{146}$  and of the fission fragments
    ($E^*_{H,L}=E_{H,L}(t_{\text{SS}})-E_{gs}(N_{H,L},Z_{H,L}$), 
    the equivalent neutron incident energy ($E_n$), the scaled initial mass moments
    $q_{20}(0)$ and $q_{30}(0)$, 
    the ``saddle-to-scission'' time $t_{SS}$, TKE evaluated as in Ref.~\cite{NuclPhys_2006r}, TKE, atomic 
    ($A_{L}^{syst}$), neutron ($N_{L}^{syst}$) and proton ($Z_{L}^{syst}$) extracted from data~\cite{PRC_2013r5}  
    using Wahl's charge systematics~\cite{Wahl_2002} and the corresponding numbers obtained in simulations,  
    and the number of post-scission neutrons  for the heavy and light fragments $(\nu_{H,L})$, estimated using a 
    Hauser-Feshbach approach and experimental neutron separation energies~\cite{PhysRev_1952r, 
    MCHF, RIPL3}. Units are MeV, fm$^2$, fm$^3$, fm/c were appropriate}

\begin{tabular}{r r r  r r r  r r r  r r r  r r r  rrr r}
\hline\hline
S\# & $\eta$ & $E^*$ & $E_n$ &  $q_{zz}$  & $q_{zzz}$ & $t_{SS}$ & TKE$^{syst}$  &  TKE & $A_L^{syst}$& $A_L$ & $N_L^{syst}$ & $ N_L$   & $Z_L^{syst}$ & $ Z_L$ &  $E^*_H $ & $ E^*_L $ & $\nu_H$ & $\nu_L$ \\
\hline 
S1   & 0.75 & 8.05  &  1.52  & 1.78 & -0.742 & 14,419   & 177.27 & 182  & 100.55  & 104.0 & 61.10  & 62.8   & 39.45    & 41.2  &  5.26   & 17.78    &  0   & 1.9  \\
S2  & 0.5   & 7.91  &  1.38  &  1.78 & -0.737 &  4,360    & 177.32 & 183 & 100.56  & 106.3 & 60.78  & 64.0   & 39.78    & 42.3  &  9.94   & 11.57    &  1   & 1 \\
S3 & 0      & 8.08  &  1.55  &  1.78  & -0.737 & 14,010   & 177.26 & 180 & 100.55  & 105.5 &  60.69 & 63.6   & 39.81    & 41.9  & 3.35    & 29.73    &  0   & 2.9 \\
S4 & 0      & 6.17  & -0.36 &  2.05   & -0.956 & 12,751   & 177.92 & 181 &             & 103.9 &            & 62.6   &              & 41.3  & 7.85    & 9.59      &  1   & 1    \\  
\hline\hline
\end{tabular}

\end{table*}

In all simulations performed by us so far the heavy fragment emerges
basically spherical and with rather small excitation energy, while the
light fragment is highly deformed and also has a higher excitation
energy. Consequently, the excitation energy of the fragments does not
follow from thermal equilibrium, as often has been assumed in the past
in phenomenological studies, see discussion in
Refs.~\cite{PhysRevLett_2010r5, PhysRevC_2011r4, PhysRevC_2011r5}, or
as a Langevin approach (which implies thermal equilibrium throughout
the entire system) might suggest.  The heavy fragment has neutron and
proton numbers very close to magic numbers and naturally very weak
pairing field as well. The large deformation energy of the light
fragment is eventually converted into a significant amount of internal
excitation energy, which is released by neutron emission and gamma
rays. The fact that the excitation energy of the heavy fragment is
significantly smaller than the excitation energy of the light fragment
correlates with the fact that the heavy fragment emerges as an almost
magic nucleus with strong shell effects~\cite{PhysRevLett_2010r5,
  PhysRevC_2011r4, PhysRevC_2011r5}.  We did not observe any
significant neutron emission at scission, conclusion confirmed
by the density profiles and the current flow we observe, see movies~\cite{movies}.  The
total kinetic energy (TKE) of the fission fragments is determined
predominantly by the elongation of the fission system at
scission~\cite{PhysRep_1990r}. In order to extract the TKE 
and the fragment excitation energies we have
assumed that after scission the internal excitation energies
do not change. When compared to existing evaluated experimental
data~\cite{NuclPhys_2006r} in the case of $^{239}$Pu(n,f) the
systematics, which follow the trend
$
{\rm TKE} = 177.80 - 0.3489 E_n\approx 177.3 {\rm \; for \; S1-S3}
\; [{\rm in} \; {\rm MeV}], 
$
we note that our estimated TKEs slightly overestimate the observed
values by at most $\approx 3\%$, see Table~\ref{table:A}.
This is indicative of the fact that in our simulations the system scissions a bit too early. 
The fission fragment mass and charge 
can be extracted from data~\cite{PRC_2013r5} (which have a resolution of
about 4-5 mass units), see Table~\ref{table:A}.
The evaluated average number of emitted
neutrons~\cite{NuclPhys_2006r} in this case is close to 3, see Ref.~\cite{ENDF}, which is
higher than the values we estimate, see Table~\ref{table:A}. If the
system would scission at a larger elongation, the light fragment would
emerge with more excitation energy and the number of emitted neutrons
would be larger.

Apart from the fact that a heavy nucleus fissions without any
restrictions on the nuclear shape, TDSLDA supplies another additional
big surprise. The time it takes a nucleus to descend from the saddle
to the scission configuration is very long. A hydrodynamic
approach~\cite{arxiv_2015nedf} and the Langevin dynamics with various
types of viscosities~\cite{PhysRevC_1976r, PhysRevC_1978r}, along with
approximate TD meanfield treatments lead to time scales of about 1000
fm/c or less. TDSLDA however, which incorporates naturally one-body
dissipation, both wall and window mechanisms~\cite{AnnPhys_1978,
  AnnPhys_1980}, points to time scales an order of magnitude larger
than predicted in the literature. The nuclear system superficially
behaves like an extremely viscous system, but the collective motion at
the same time is not overdamped. There is a significant amount of
collective flow, which is not dissipated and transformed into
heat. The slide of the nucleus down from the saddle to the scission is
not a monotonic one, but it is accompanied by a significant amount of
collective shape and pairing field excitations in ``transverse
directions," see Fig. \ref{fig:Delta}.  The long
``saddle-to-scission'' time $t_{SS}$ can be attributed in part to the weak
proton pairing gap in the starting configuration.  In cases where the
system starts initially with a relatively weak pairing proton gap,
during the slide the proton pairing gap shows large temporal
and spatial fluctuations~\cite{movies}. In contradistinction in the 
TDHF+BCS approximation, spatial fluctuations are 
absent, the phase of the pairing field can be eliminated by a trivial 
gauge transformation, and fission does not happen 
without boost from configurations near the outer 
saddle~\cite{PhysRevC_2015r1, PhysRevC_2015r2}.
These large pairing gap fluctuations facilitate the shape evolution
and the formation of the neck and the eventual scission of the
nucleus.  The two-body dissipation effects might affect these
conclusions.  A similar increase of the evolution time was
demonstrated by Caldeira and Leggett~\cite{Physica_1983,
  AnnPhys_1983}, when coupling a simple quantum system with a
``thermal bath;" see also Refs.~\cite{AB&DK_1995,  NuclPhys_1989r1} 
and the "bearing balls video" (Drude model for electrons)~\cite{movies}.
Phenomenologicaly~\cite{PhysRevLett_2011r,PhysRevC_2011r}
the fission fragments
distribution is reconstructed from (over)damped dynamics, thus on very
long time scales, which superficially is in agreement
with a time averaging of our microscopic dynamics and
with the apparent significantly reduced role of collective inertia in
the dynamics in a reduced collective space.

We have explored only axial symmetric
configurations with broken left-right/parity symmetry 
($q_{zzz}\neq 0$). Most authors agree that axial symmetry 
is hardly ever broken beyond the outer saddle. The system spontaneously
has chosen such an initial deformation after we have imposed a slight
pinch slightly off the middle of the mother nucleus. There is
collective matter flow from one side to the other of the nucleus
before scission and the system determines dynamically its final
fragment sizes, see movie~\cite{movies}.  This is indicative of the
character of the potential energy surface, which shows softness in
this collective variable, which was observed in previous
studies~\cite{PhysRevC_2009r, PhysRevC_2011r1, PhysRevC_2014r1,
  PhysRevC_2014r2}.  The axial symmetry can be broken either spontaneously 
  initially (not observed by us) or by quantum fluctuations (not studied here) during the
evolution.  

The quality of the agreement with experimental observations surprised
us in its accuracy, since we have made no effort to reproduce any
measured data. We have merely used a rather randomly chosen NEDF, with
rather decent properties, but far from perfect. However, since this
NEDF encodes reasonably well gross nuclear properties it does not come
as a great surprise that gross properties of nuclear fission emerge so
close to what one might have hoped for. Clearly, the detais of the
energy density functional at large deformations and the details of the
pairing interaction will have to be pinpointed with greater
accuracy. Induced nuclear fission 
presents us with a 
unique opportunity, in this respect, as in the study of the ground and weakly excited
states, and even in the case of spontaneous
fission~\cite{PhysRevC_2014r1}, one can explore only rather
small nuclear deformations.  The nature of the dynamics of a
fissioning nucleus appears quite surprising, the overall rolling down
the hill is much slower than ever expected, but not
because of a particularly large viscosity. Rather, a large number of
CDOF are excited, both shape and pairing modes, 
clearly demonstrated in the real-time movies \cite{movies}.
The strong energy exchange between a large number of CDOF appears to be at the root
of the slowness of this unexpected dynamics. There are
experimental indications that fission times can be extremely 
long~\cite{PhysRevLett_2008r4, PhysRevLett_2011r4, EPJ_2013r4}.

Even though in this first study of its kind we did not obtain a
perfect agreement with experiment, our results clearly demonstrate
that rather complex calculations of the real-time fission dynamics
without any restrictions are feasible and further improvements in the
quality of the NEDF, and especially in its dynamic properties, can
lead to a theoretical microscopic framework with great predictive
power, where experiments are not feasible, particularly 
in astrophysical environments. Extension of the present approach to
two-body observables (fission fragment mass, charge, angular 
momenta, and excitation energies
distribution widths) are rather straightforward to
implement~\cite{AnnPhys_1985r1, AnnPhys_1988r,
  PhysRevLett_2011r1} and eventually more detailed information could
be inferred by introducing the stochasticity of the
meanfield~\cite{JPhysG_2010, arxiv:2015r5}.

 
We have greatly benefitted from numerous discussions with G.F. Bertsch
and over the years from interactions with A.B. Migdal, P. M\"oller,
J. Randrup, A. Sierk, W.J. Swiatecki, and P. Talou.  We thank J. Wells
for his continuous support. We thank G.F. Bertsch, A. Sierk, and
P. Talou for a careful reading of the manuscript and suggestions as well as 
the referees for very thoughtful suggestions.
This work was supported in part by U.S. Department of Energy (DOE)
Grant No. DE-FG02-97ER41014, the Polish National Science Center (NCN)
under Contracts No. UMO-2013/08/A/ST3/00708 and
No. UMO-2012/07/B/ST2/03907.  IS gratefully acknowledges partial
support of the U.S. Department of Energy through an Early Career Award
of the LANL/LDRD Program.  Calculations have been performed at the
OLCF Titan~\cite{cray} and at NERSC Edison.  This research used
resources of the Oak Ridge Leadership Computing Facility, which is a
DOE Office of Science User Facility supported under Contract
DE-AC05-00OR22725 and 
used resources of the National Energy Research
Scientific computing Center, which is supported by the Office of
Science of the U.S. Department of Energy under Contract
No. DE-AC02-05CH11231 The contribution of each one of the authors has
been significant and the order of the names is alphabetical.


\end{document}